\documentclass{article}
\usepackage{graphicx}  
\usepackage{amsmath}   
\usepackage[compress]{cite}
\usepackage{amssymb}   
\usepackage{bm} 
\usepackage{dcolumn}
\usepackage{color}
\usepackage{mathrsfs}
\usepackage{gensymb}
\usepackage{float}
\usepackage{amsfonts}
\usepackage{inputenc}
\usepackage{varioref}
\usepackage{textcomp}
\usepackage[caption=false]{subfig}
\RequirePackage[colorlinks,citecolor=blue,urlcolor=magenta,linkcolor=blue]{hyperref}
\allowdisplaybreaks
\def\gr{general relativity}

\def\KN{Kerr-Newmann }
\labelformat{section}{Section #1} 
\labelformat{subsection}{Section #1} 
\labelformat{subsubsection}{Section #1}
\labelformat{subsubsubsection}{Section #1}
\labelformat{equation}{Eq.~(#1)} 
\labelformat{figure}{Fig.~#1} 
\labelformat{subfigure}{Fig.~\thefigure#1} 
\labelformat{table}{Tab.~#1} 
\labelformat{appendix}{Appendix #1}
\title{Silhouette of M87*: A new window to peek into the world of hidden dimensions}
\author{Indrani Banerjee\footnote{tpib@iacs.res.in}~$^{1}$, Sumanta Chakraborty\footnote{sumantac.physics@gmail.com}~$^{1,2}$ and Soumitra SenGupta\footnote{tpssg@iacs.res.in}~$^{1}$\\
{$^{1}$\small{School of Physical Sciences, Indian Association for the Cultivation of Science, Kolkata-700032, India}}\\
{$^{2}$\small{School of Mathematical and Computational Sciences, Indian Association for the Cultivation of Science}}\\
{\small{Kolkata-700032, India}}}
\begin{document}

\maketitle
\begin{abstract}
The recent observation of the shadow of the supermassive black hole M87*, located at the centre of the M87 galaxy, by the Event Horizon Telescope collaboration has opened up a new window to probe the strong gravity regime. In this letter, we explicitly demonstrate the consequences of this observation on brane world black holes, characterized by existence of a negative tidal charge. Our results based on three observables associated with the shadow, namely, angular diameter, deviation from circularity and axis ratio reveal that the existence of a negative tidal charge is more favoured, possibly marking a deviation from \gr.
\end{abstract}
\section{Introduction}

Gravitational interaction in the strong field regime, e.g., near the horizon of a black hole (BH) is supposed to provide a wealth of information regarding the nature of gravity at a fundamental level. Until recently, \emph{direct} observations probing the near horizon regime of a BH geometry were not available. With the two successive ground breaking discoveries, namely the gravitational wave measurements from the collision of binary BHs and neutron stars \cite{Abbott:2016blz,TheLIGOScientific:2017qsa} and imaging the shadow of the supermassive BH at the centre of the M87 galaxy \cite{Fish:2016jil,Akiyama:2019cqa,Akiyama:2019brx,Akiyama:2019sww,Akiyama:2019bqs,Akiyama:2019fyp,Akiyama:2019eap}, the prospects of understanding the nature of strong gravity has largely enhanced. 
 
Till date, general relativity is the most successful candidate for explaining gravitational interactions at all length scales. Although predictions from \gr\ are in well-agreement with the observations related to gravitational waves and BH shadow, there are several reasons to look for theories beyond \gr. This includes existence of spacetime singularities, loss of predictability of Einstein's equations beyond Cauchy horizons \cite{Penrose:1964wq,Hawking:1976ra,Christodoulou:1991yfa,Dafermos:2003wr,Cardoso:2017soq,Rahman:2018oso} and the necessity to invoke exotic forms of matter and energy like the dark matter and dark energy \cite{Bekenstein:1984tv,Milgrom:2003ui,Perlmutter:1998np,Riess:1998cb}. Further, even in the strong field regime, the observational avenues have not been properly explored, e.g., the quasi-normal modes in the gravitational wave signal originating from the collision of two black holes has not been probed to a great accuracy due to low signal to noise ratio \cite{Flanagan:1997sx}. Similarly, the measurement of the shadow of the supermassive black hole M87* also has its own share of limitations \cite{Akiyama:2019cqa,Akiyama:2019brx,Akiyama:2019sww,Akiyama:2019bqs,Akiyama:2019fyp}. Thus it is more than important at the present epoch to see what these observations can tell us about certain theories of gravity beyond \gr. 

Among the various alternatives to \gr, in this work we concentrate on the modification due to the presence of extra spatial dimensions \cite{Maartens:2003tw,Kanti:2004nr,PerezLorenzana:2005iv,Csaki:2004ay}. There are plenty of motivations for introducing such extra dimensions, starting from the unification of electromagnetism and gravity by Kaluza and Klein \cite{Overduin:1998pn} to resolution of the gauge hierarchy problem \cite{ArkaniHamed:1998rs,Randall:1999ee}. The presence of  extra dimensions modify the gravitational dynamics on the four-dimensional time-like hypersurface we live in (the 3-brane). This in turn is expected to leave some observable imprint on the strong field tests of gravitational interaction. It turns out that BH solutions associated with such effective gravitational field equations resemble the \KN metric \cite{Shiromizu:1999wj,Dadhich:2000am,Harko:2004ui,Carames:2012gr,Chakraborty:2016gpg,Chakraborty:2015taq} where the tidal charge can be negative unlike \gr\ \cite{Dadhich:2000am,Harko:2004ui,Chakraborty:2014xla,Aliev:2005bi}. Implications of such higher dimensional gravity theories have already been discussed in detail in the context of gravitational wave observations \cite{Chakravarti:2019aup,Chakravarti:2018vlt,Chakraborty:2017qve,Visinelli:2017bny}, but a similar discussion for BH shadow from M87* is lacking (see, however, \cite{Vagnozzi:2019apd}). To fill this gap, in this work we will discuss the implications of the observed BH shadow from M87* on theories with extra spatial dimensions. 

Possible constraints on extra dimensions from weak field tests of gravity \cite{Bhattacharya:2016naa}, electromagnetic observations of quasars \cite{Banerjee:2017hzw,Banerjee:2019sae,Banerjee:2019cjk}, collision of binary black holes and neutron stars \cite{Chakravarti:2019aup,Chakravarti:2018vlt,Chakraborty:2017qve,Visinelli:2017bny} and calculation of quadrupole moment \cite{Vagnozzi:2019apd} have already been reported. In this paper we investigate the shadow of such braneworld black holes in light of the observed silhouette of M87* by the Event Horizon Telescope collaboration. This enables us to comment on the signature of the tidal charge from one of the direct strong field tests of gravity. 

The letter is organized as follows: We start by summarizing the nature of the BH shadow in the presence of an extra spatial dimension, following which, we have defined various observables characterizing the BH shadow. These observables have been used later on, in the context of the observed shadow of M87*, to present possible constraints on the tidal charge parameter inherited from the extra dimension. Finally we conclude with a discussion of our results.  

\emph{Notations and Conventions:} We use mostly positive metric signature convention and set the fundamental constants $G$ and $c$ to unity. The Greek indices $(\mu,\nu,\cdots)$ are used to denote brane coordinates, while the bulk coordinates are represented by the capitalized latin indices $(A, B, \cdots)$. 
\section{Black hole shadow on the brane}

In this section we discuss the framework necessary for the computation of BH shadow in the effective 4-d theory, originating from a 5-d spacetime (bulk). We assume that the gravitational action in the bulk consists of only the Einstein-Hilbert term such that the gravitational field equations are represented by the Einstein's equations. These equations when projected on the 3-brane, yields the effective gravitational field equations, different from the Einstein's equations, having the form,
\begin{align}
~^{(4)}G_{\mu \nu}+E_{\mu \nu}=8\pi G_{4} \left\{T_{\mu \nu}+\frac{6}{\lambda_{\rm b}}\tau_{\mu \nu}\right\}~.
\end{align}
Here, $~^{(4)}G_{\mu \nu}$ is the Einstein tensor constructed out of brane geometry alone and $E_{\mu \nu}=W_{PAQB}e^{P}_{\mu}n^{A}e^{Q}_{\nu}n^{B}$, represents the non-local effects of the bulk through the bulk Weyl tensor $W_{ABCD}$, the projectors $e^A_\alpha$ and the normalized normals to the brane $n^C$. The matter sector, on the other hand, has the additional term $\tau_{\mu \nu}$ containing quadratic corrections of brane energy momentum tensor $T_{\mu \nu}$. The coupling coefficient $\lambda_{\rm b}$ appearing in the above equation is known as the brane tension and is another characteristic parameter from the extra dimensions \cite{Shiromizu:1999wj,Dadhich:2000am,Germani:2001du,Maartens:2003tw}. Therefore, a brane observer perceives the effect of the bulk through the aforementioned modifications in the gravitational field equations. It is possible to write down an axially symmetric BH solution on the brane by solving the above effective gravitational field equations in vacuum, which takes the form,
\begin{align}\label{papereq1}
ds^{2}&=-\bigg(\frac{\Delta-a^{2}\sin^{2}\theta}{\rho^{2}}\bigg)dt^2-\frac{2a\sin^2\theta\left(r^{2}+a^{2}-\Delta\right)}{\rho^{2}} dt d\phi 
+ \frac{\rho^{2}}{\Delta}dr^2+\rho^{2} d\theta^2 
\nonumber
\\
&+\bigg\{r^2 + a^2 +\frac{a^2 \sin^2\theta\left(r^{2}+a^{2}-\Delta\right)}{\rho^{2}}\bigg\}\sin^2\theta d\phi^2~,
\end{align}
where $\rho^{2}=r^2+a^2 \cos^2\theta$ and $\Delta=r^2 -2Mr+a^2+4M^{2}q$ \cite{Aliev:2009cg}. Here $M$ is the mass of the black hole, $J=aM$ is the black hole angular momentum and $q$ is the dimensionless tidal charge parameter inherited from higher dimensions. Note that, the charge parameter $q$ can assume both positive and negative values. For positive values of $q$, \ref{papereq1} appears identical to a \KN BH, while the case with negative $q$ has no analogue in GR and thus provides a true signature of the additional spatial dimensions \cite{Aliev:2005bi,Aliev:2009cg,Neves:2012it}. We would like to reiterate that this solution is derived with $T_{\mu \nu}=0$, i.e., in vacuum and the tidal charge term is inherited from the non-local contributions of the bulk spacetime, which has no analogue in \gr.

We now discuss the effect of this tidal charge on the BH shadow. The shadow of a BH consists of a dark region in the sky of the observer (us) located at a large distance from the BH. The shadow is formed when photons with small impact parameter originating from a source of electromagnetic radiation behind the BH gets trapped by its gravity \cite{Hioki:2009na,Abdujabbarov:2015xqa,Abdujabbarov:2016hnw,Mishra:2019trb,Gralla:2019xty,Bambi:2019tjh,Cunha:2018acu,Tian:2019yhn} and hence fail to reach the observer. The nature of the BH shadow for Kerr as well as \KN BHs have been extensively studied in \cite{Cunha:2018acu,Vries_1999}, while for braneworld BHs one is referred to \cite{Eiroa:2017uuq,Amarilla:2011fx,Abdujabbarov:2017pfw}. Here we briefly summarize the procedure to construct the BH shadow in a given spacetime.

\begin{figure}
\begin{center}

\subfloat[The above figure demonstrates the structure of the shadow for a non-rotating black hole for various tidal charge parameters.]{\includegraphics[scale=0.7]{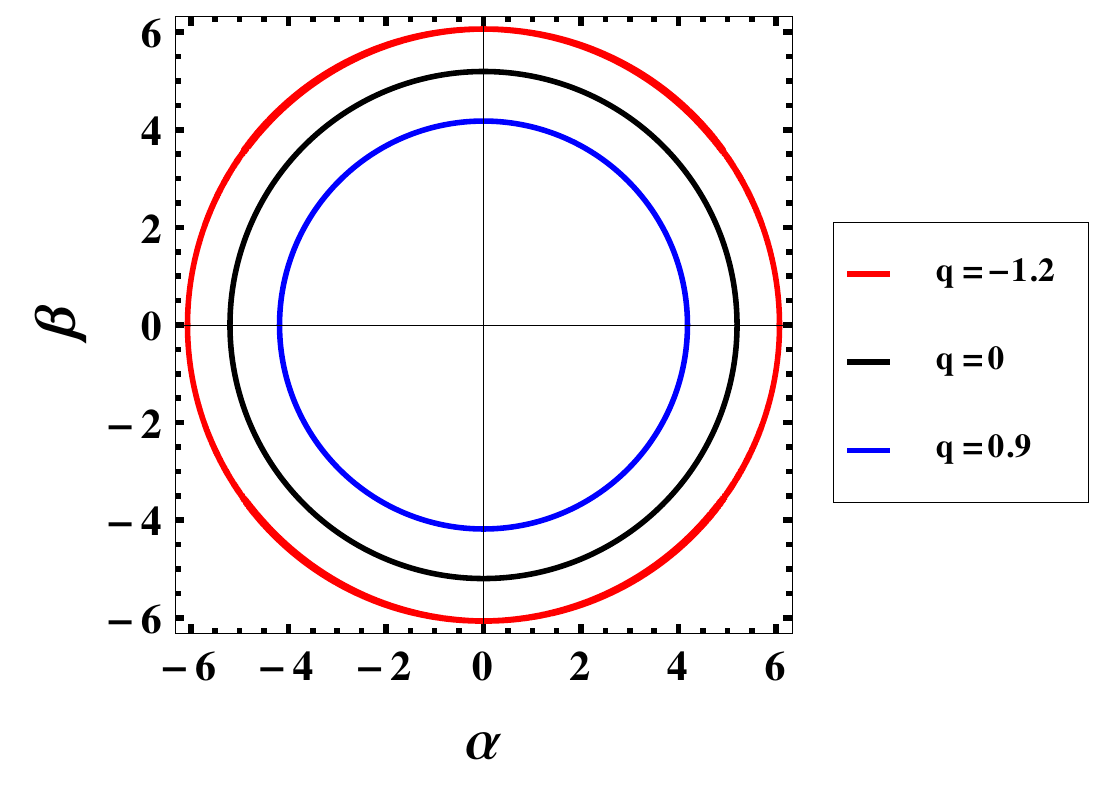}}
\hfill
\subfloat[The structure of the shadow for a rotating black hole with $a=0.8$ and inclination angle $0^{\circ}$ has been presented for various choices of the tidal charge parameter.]
{\includegraphics[scale=0.7]{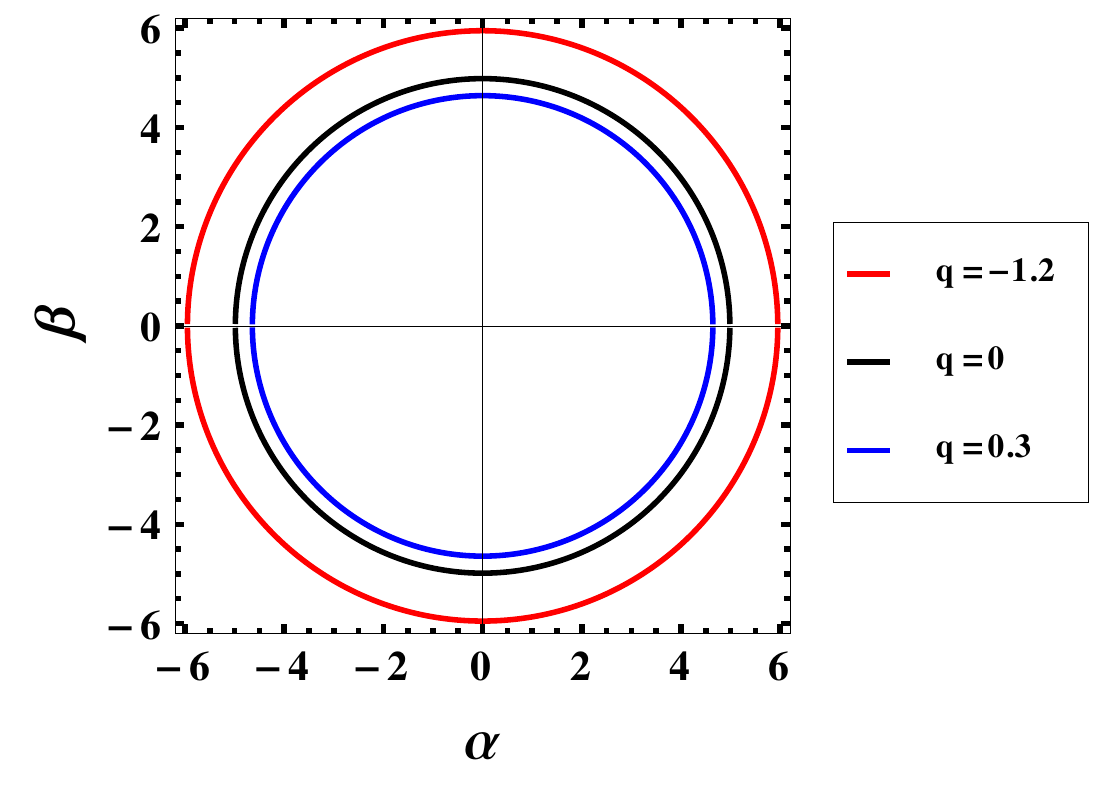}}
\\
\subfloat[The above figure demonstrates how the structure of the black hole shadow changes as the tidal charge parameters changes from positive to negative for a black hole with $a=0.8$ and inclination angle $60^{\circ}$.]{\includegraphics[scale=0.7]{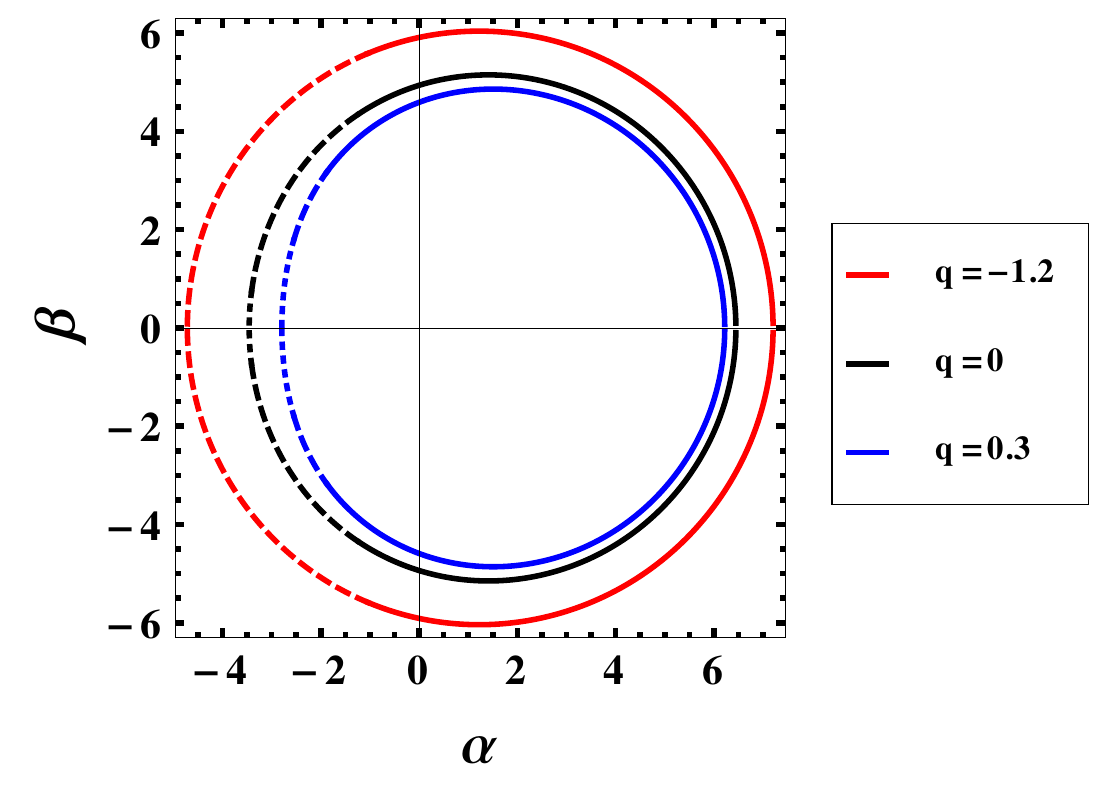}}
\hfill
\subfloat[The structure of black hole shadow with $a=0.8$ and inclination angle $90^{\circ}$ has been shown for different tidal charge parameters.]
{\includegraphics[scale=0.7]{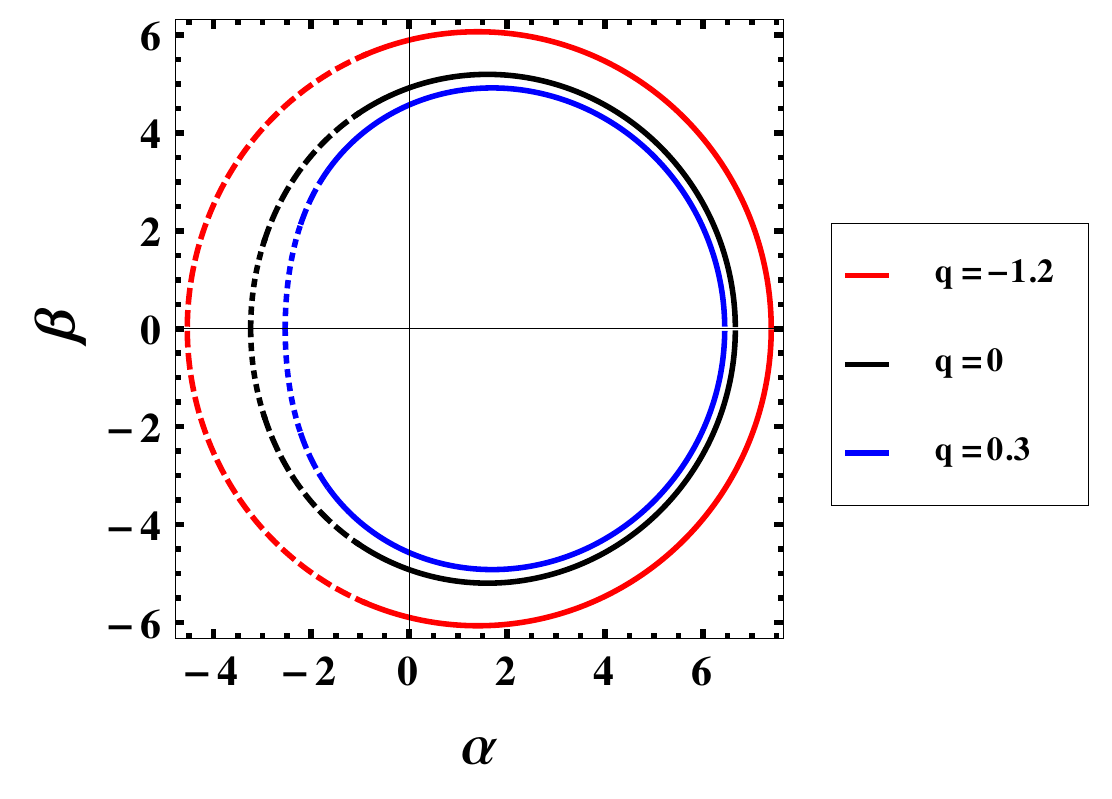}}
\caption{Shadow structure for a black hole has been plotted in the plane of celestial coordinates $(\alpha,\beta)$ for various choices of the black hole hairs, namely the rotation parameter $a$ and tidal charge parameter $q$ along with the inclination angle $\theta_{0}$.}\label{Fig_Shadow_Theory}
\end{center}
\end{figure}

The starting point is the geodesic equations of a massless particle in the spacetime metric described by \ref{papereq1}. It is well known that the geodesic equations are separable due to the presence of a Killing tensor. Consequently, there are three constants of motion, the energy $E$, angular momentum $L$ and the Carter constant $K$ \cite{Carter:1968rr}. Using these three constants of motion one can define two possible impact parameters $\xi\equiv (L/E)$ and $\eta\equiv(K/E^{2})$, which denote the perpendicular distance from the axis of rotation and the equatorial plane, respectively. Note that for non-zero Carter constant the motion of the photon is not confined to the equatorial plane, which endows its orbital motion a much rich structure \cite{Teo2003}. 

To construct the BH shadow in the observer's sky, it is customary to consider the observer to be at a large distance $r_{0}$ from the BH with an inclination angle $\theta_{0}$ from the rotation axis. This enables one to define two celestial coordinates $\alpha$ and $\beta$, dependent on the radius of the photon circular orbit $r_{\rm ph}$ through the impact parameters $\xi$ and $\eta$ \cite{Bardeen:1973tla,Vries_1999} as well as the inclination angle $\theta_{0}$ as,
\begin{align}
\alpha&=\lim_{r_{0}\rightarrow \infty}\left(-r_{0}^{2}\sin \theta_{0}\frac{d\phi}{dr}\right)=-\xi~ \textrm{cosec}\theta_{0}~;
\\
\beta&=\lim_{r_{0}\rightarrow \infty}\left(r_{0}^{2}\frac{d\theta}{dr}\right)=\pm \sqrt{\eta+a^{2}\cos^{2}\theta_{0}-\xi^{2}\cot^{2}\theta_{0}}~.
\end{align}
Eliminating $r_{\rm ph}$ from $\alpha$ and $\beta$ one obtains the contour of the shadow in the observer's sky, dependent on the inclination angle $\theta_{0}$ \cite{Cunha:2018acu}. The shape and size of the shadow depends primarily on the background spacetime, e.g., the size of the shadow directly scales with the BH mass. In particular, it turns out that presence of a negative tidal charge parameter enhances the size of the shadow compared to \gr. To illustrate the dependence of the black hole shadow on black hole hairs in a clear manner, we have presented the variation of the contours of the black hole shadow with tidal charge $q$, rotation parameter $a$ and inclination angle $\theta_0$ in \ref{Fig_Shadow_Theory}. Note that if an observer is at zero inclination angle with respect to the black hole rotation axis, the shadow will always be circular, although its radius will depend on the choice of the rotation parameter.

\section{Observables associated with BH shadow}

Having discussed the basic properties of the braneworld BH shadow, we now describe the observables which can be used to test the hairs of the BH \cite{Kumar:2018ple}. This in turn will enable us to understand whether effects beyond GR are present.

\begin{figure}[htb]
\begin{center}
\includegraphics[scale=0.3]{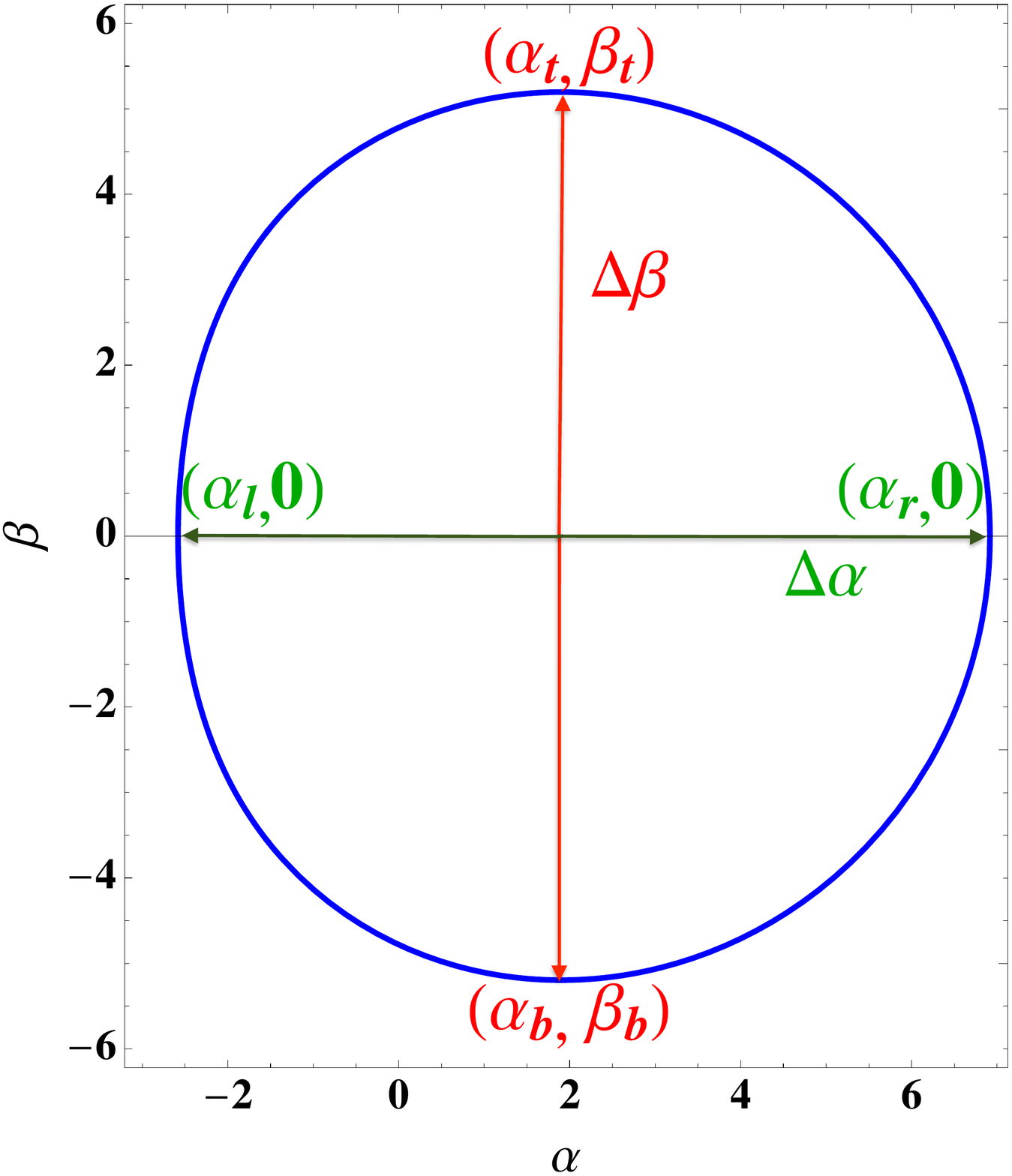}
\caption{The schematic structure of BH shadow in the $(\alpha,\beta)$ plane with non-zero $a$ and $\theta_0$ has been presented. The two diameters $\Delta \beta$ and $\Delta \alpha$, along with the relevant points on the black hole shadow are also depicted. These diameters are used to construct various observables associated with the black hole shadow.}\label{Fig_Schematic}
\end{center}
\end{figure}

The most important observable associated with the BH shadow corresponds to the shadow diameter $\Delta \beta$. Since the shadow is not circular in general (see \ref{Fig_Schematic}), there can be two possible diameters, the major axis $\Delta \beta=\beta_{t}-\beta_{b}$ and the minor axis $\Delta \alpha=\alpha_{r}-\alpha_{l}$, which coincides with $\Delta \alpha$ if the shadow is circular. The boundary of the shadow can be considered as a curve $\beta(\alpha)$ and $\Delta \alpha$ and $\Delta \beta$ can be expressed in terms of the metric parameters and inclination angle $\theta_{0}$ \cite{Wei:2019pjf}. Note that the analysis presented here assumes a knowledge about the orientation of the $(\alpha,\beta)$ plane, which may not be true in generic situations.

\begin{figure}[H]
\begin{center}
\subfloat[Estimation of the Shadow diameter $\Delta \beta$ has been presented in the $(a,q)$ plane for an inclination angle of $0^{\circ}$.]
{\includegraphics[scale=0.5]{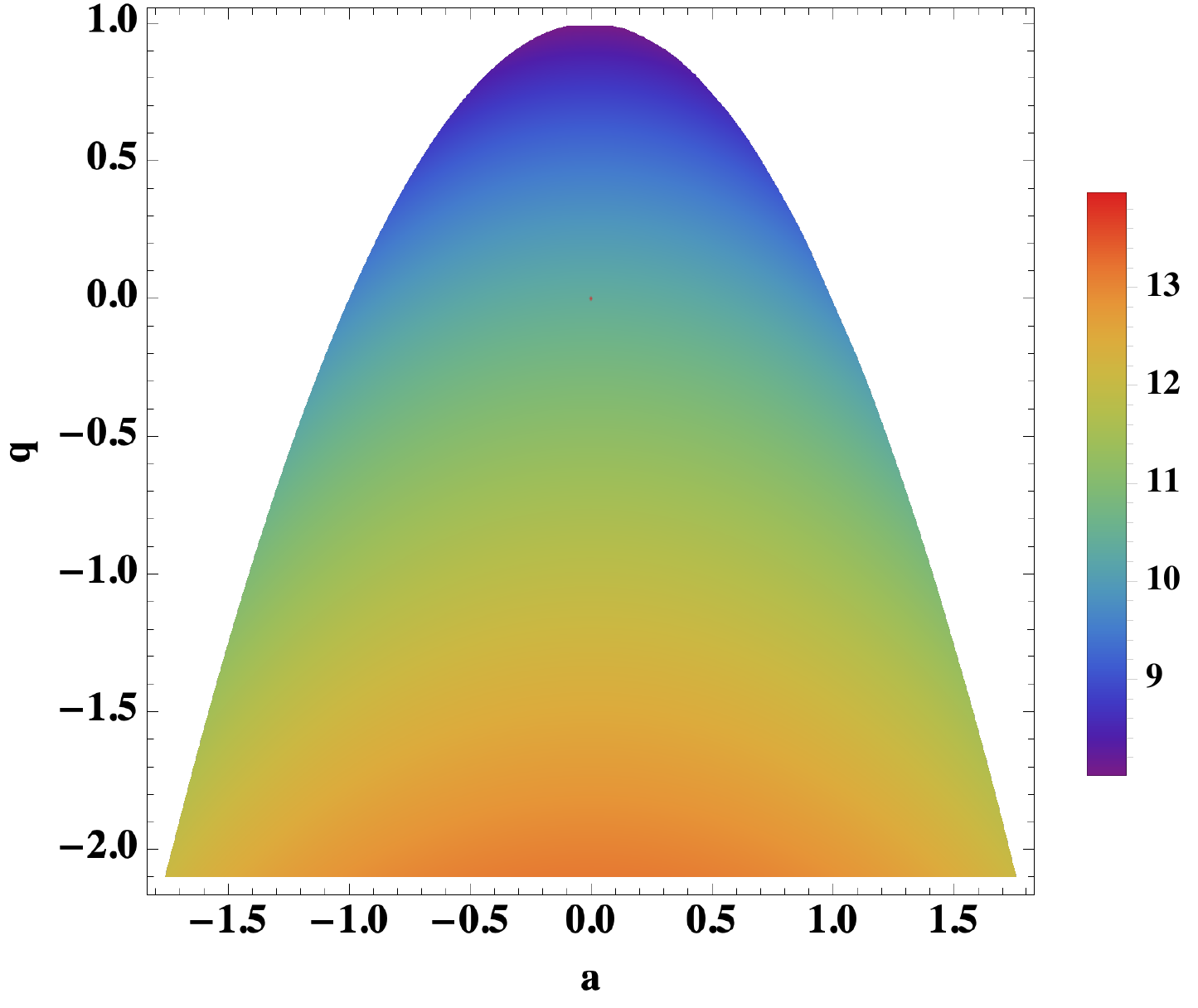}}
\hfill
\subfloat[The numerical estimates of the shadow diameter $\Delta \beta$ has been depicted for an inclination angle of $45^{\circ}$.]
{\includegraphics[scale=0.5]{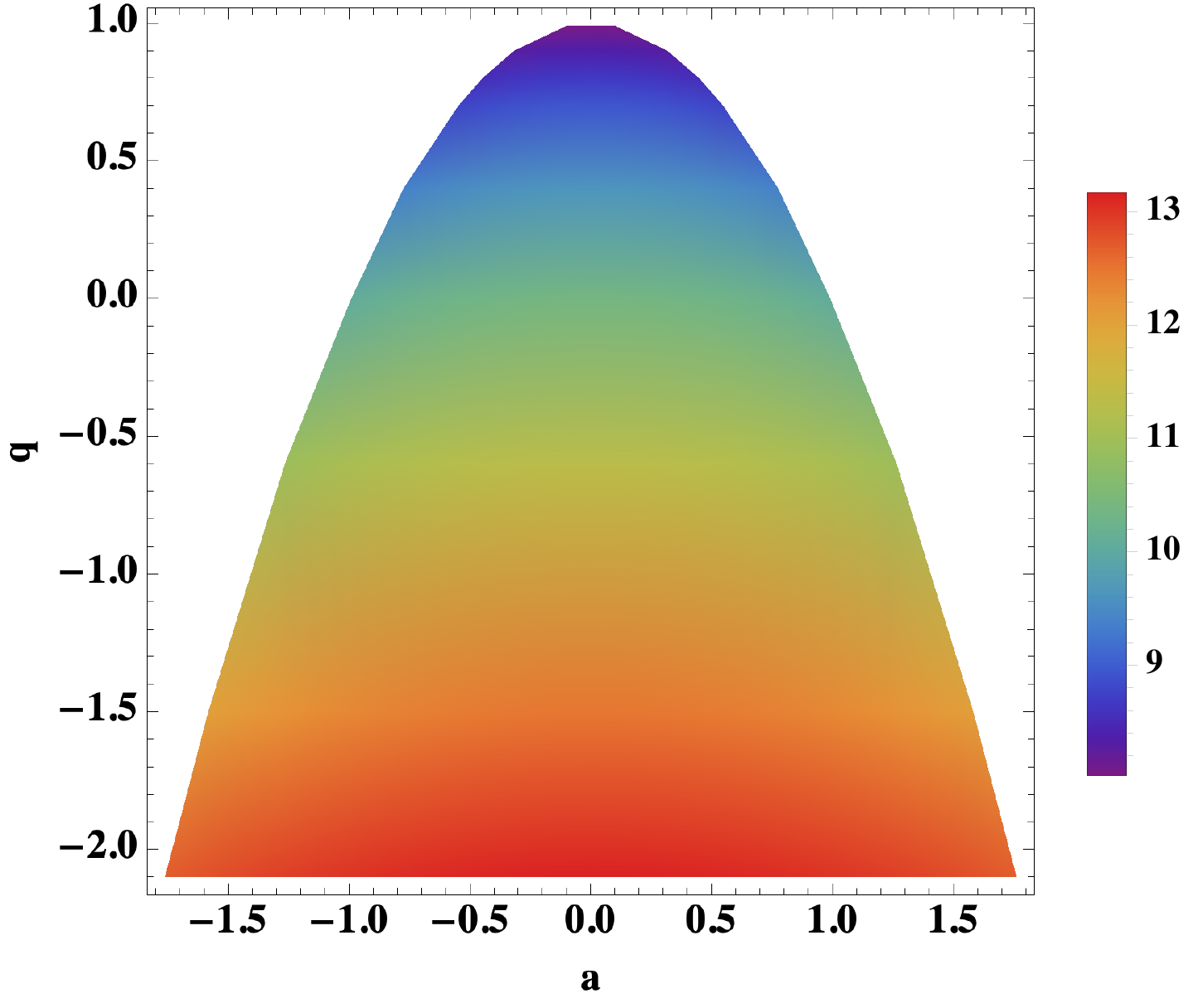}}
\\
\subfloat[The variation of the shadow diameter $\Delta \beta$ with rotation parameter $a$ has been presented for the inclination angle $90^{\circ}$.]
{\includegraphics[scale=0.8]{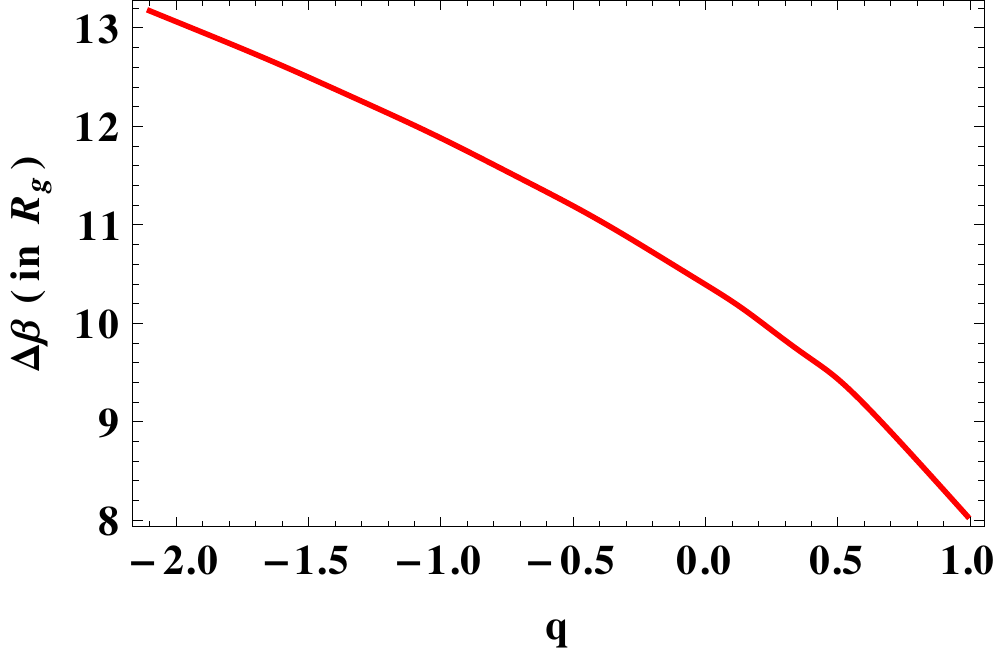}}
\caption{The shadow diameter $\Delta \beta$ has been plotted in the $(a,q)$ plane for different inclination angles.}\label{Fig_DeltaBeta_Theory}
\end{center}
\end{figure}

\begin{figure}
\begin{center}
\subfloat[$\Delta C$ has been plotted in the $(a,q)$ plane for an inclination angle of $45^{\circ}$.]
{\includegraphics[scale=0.5]{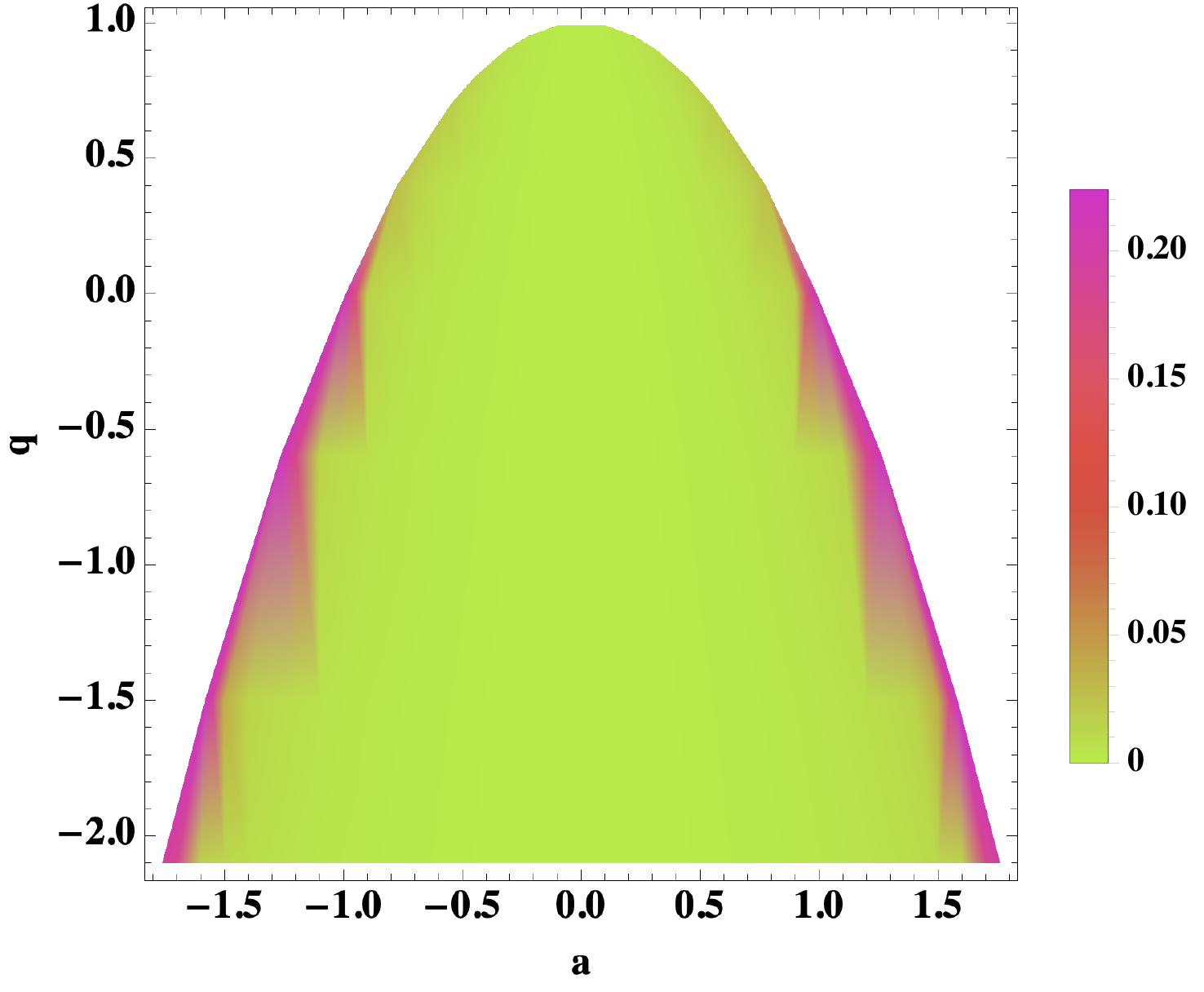}}
\hfill
\subfloat[The axis ratio $(\Delta \beta/\Delta \alpha)$ is presented in the $(a,q)$ plane for an inclination angle of $45^{\circ}$.]
{\includegraphics[scale=0.5]{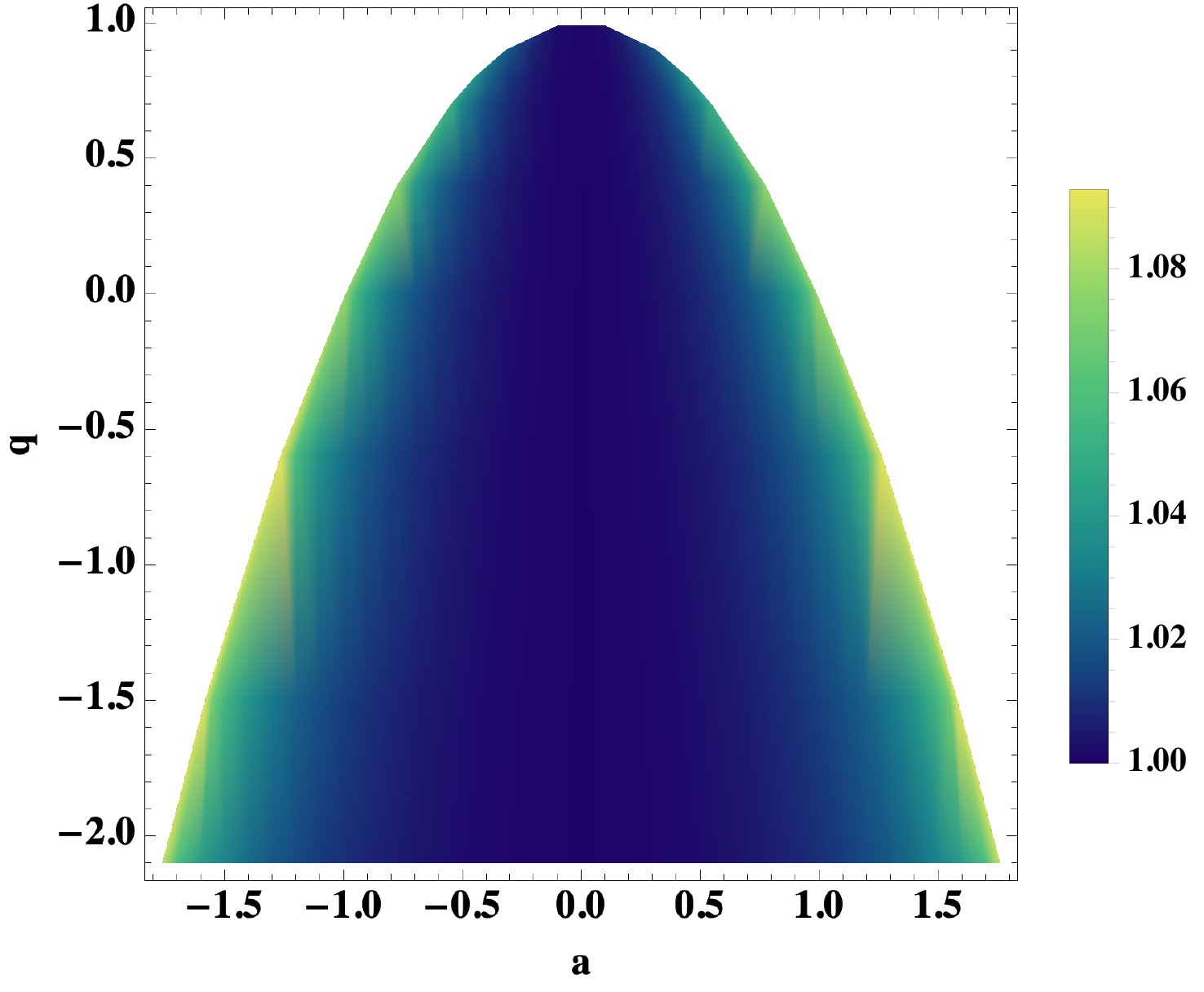}}
\\
\subfloat[Numerical estimates of the deviation from circularity, $\Delta C$ has been presented  in the $(a,q)$ plane for inclination angle of $90^{\circ}$.]
{\includegraphics[scale=0.5]{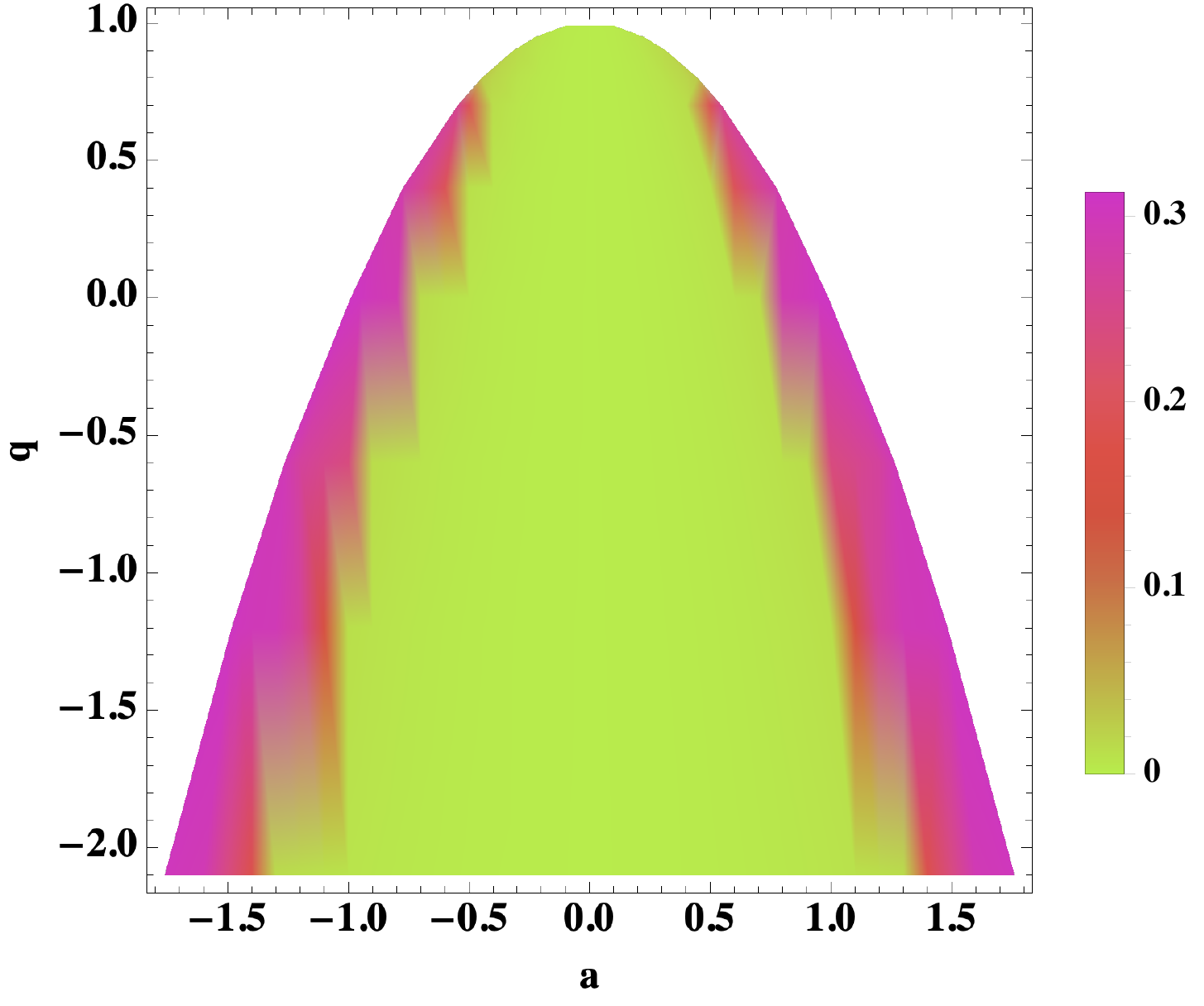}}
\hfill
\subfloat[Axis ratio $(\Delta \beta/\Delta \alpha)$ is shown in the $(a,q)$ plane for an inclination angle of $90^{\circ}$.]
{\includegraphics[scale=0.5]{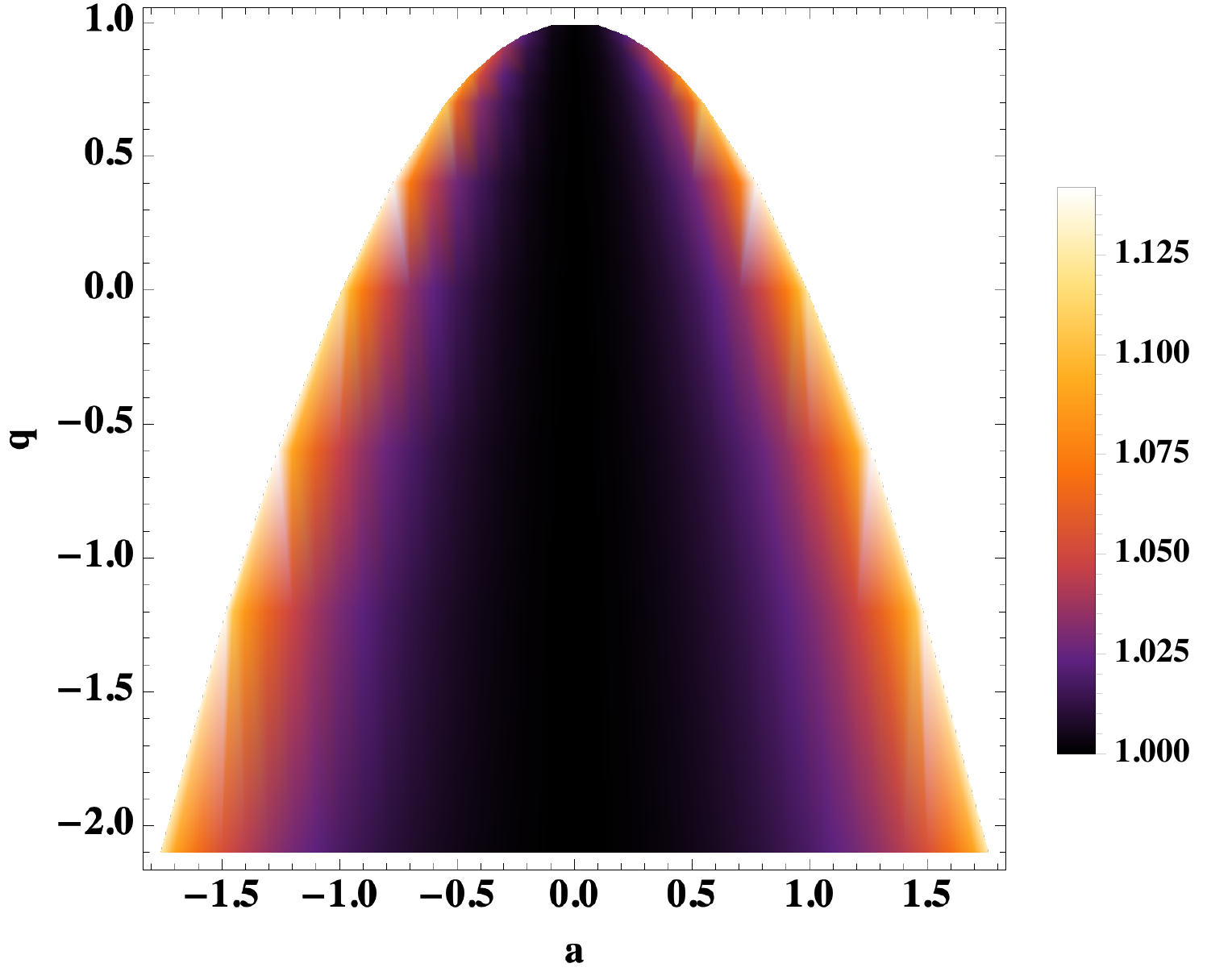}}
\caption{The two observables, namely deviation from circularity $\Delta C$ and axis ratio $\Delta A=(\Delta \beta/\Delta \alpha)$ have been plotted in the $(a,q)$ plane for different values of the inclination angle. See text for discussions.}\label{Fig_AxisRatio_DeltaC_Theory}
\end{center}
\end{figure}

Another observable comes from the fact that a rotating BH, when viewed from a non-zero inclination angle exhibits a dented circular shape. Thus deviation of the shadow from circularity $\Delta C$ is another primary observable which can be defined as (see \cite{Bambi:2019tjh}),
\begin{align}\label{papereq2}
\Delta C=\frac{1}{R_{\rm av}}\sqrt{\frac{1}{2\pi}\int_{0}^{2\pi}d\phi ~ \left(\ell(\phi)-R_{\rm av}\right)^{2}}
\end{align}
such that $\ell(\phi)=\sqrt{(\alpha(\phi)-\alpha_{c})^{2}+\beta(\phi)^{2}}$ is the distance between the geometric centre, located at $\alpha_{c}=(1/\textrm{Area})\{\int dA~\alpha\}, \beta_{c}=0$ and any point on the boundary of the shadow with azimuthal angle $\phi$. Further, the average radius $R_{\rm av}$ has the following expression \cite{Bambi:2019tjh},
\begin{align}
R_{\rm av}^{2}=\frac{1}{2\pi}\int _{0}^{2\pi} d\phi ~ \ell^{2}(\phi)
\end{align}
The celestial coordinates $\{\alpha(\phi),\beta(\phi)\}$ depends on the mass $M$, rotation parameter $a$, the tidal charge $q$ and the inclination angle $\theta_{0}$. Therefore, the radius $R_{\rm av}$ depends on these parameters as well. This in turn leads to the expression for $\Delta C$, which also depends on these parameters. Thus for a given inclination angle, the deviation from circularity will be functions of rotation parameter $a$ and tidal charge $q$. Following which we have plotted $\Delta C$ in the $(a,q)$ plane for inclination angles of $45^{\circ}$ and $90^{\circ}$ respectively in \ref{Fig_AxisRatio_DeltaC_Theory}. As the left plots presented in \ref{Fig_AxisRatio_DeltaC_Theory} demonstrate, for a smaller inclination angle, the above estimate of the deviation from circularity is very small, while it increases as the inclination angle increases. On the other hand, for $a=0$, the deviation from circularity identically vanishes, while as $a$ increases the deviation also increases. Finally, for a fixed rotation parameter $a$, as the tidal charge increases from negative to positive values, the deviation from circularity also increases. Thus if the deviation from circularity of the shadow can be explicitly measured, it will possibly lead to interesting bounds on the $(a-q)$ plane given the inclination angle.   

Another observable of similar nature correspond to the axis ratio $\Delta A\equiv \{\Delta \beta/\Delta \alpha\}$, where $\Delta \beta$ is the largest distance between two points on the boundary of the shadow, known as the shadow diameter and $\Delta \alpha$ is the extent of the shadow on the $\beta=0$ line. For non-rotating or, vanishing inclination angle these two lengths match, yielding $\Delta A=1$. On the other hand, for a rotating black hole viewed from a non-zero angle, the two lengths will not be the same and hence $\Delta A$ will be different from identity. Thus if one plots $\Delta A$ in the $(a,q)$ plane for a rotating black hole with non-vanishing inclination angle it will also depict a measure of deviation from circularity. Following which we have plotted the behaviour of axis ratio $\Delta A$ in the $(a,q)$ plane with different inclination angles in \ref{Fig_AxisRatio_DeltaC_Theory}. As evident from the plots in the right column, with increase in the inclination angle the axis ratio also increases, which is consistent with the enhancement of the deviation from circularity. Thus our theoretical expectations are borne out in \ref{Fig_AxisRatio_DeltaC_Theory}, which when applied to the real data from black hole shadow measurement, may shed some light on the nature of gravity in the strong field regime.
\section{Shadow of M87* and extra dimensions}

In this section, we use the results from the EHT to discern the observationally favoured values of $a$ and $q$. The observed shadow of M87* has an angular diameter of $(42\pm 3)~\mu\textrm{as}$, while it exhibits an axis ratio $<4/3$ and the deviation from circularity is reportedly less than 10\% \cite{Akiyama:2019cqa,Akiyama:2019fyp,Akiyama:2019eap}. To compare our results with the EHT observation we compute the angular diameter of the shadow, which apart from $a$, $q$ and $\theta_0$ depends largely on the mass $M$ and the distance $D$ of the BH. An increase in mass and a decrease in distance leads to an enhancement of the angular diameter. The inclination angle is taken to be $17^{\circ}$, which the jet axis makes to the line of sight \cite{Akiyama:2019cqa,Akiyama:2019fyp,Akiyama:2019eap}. The distance $D$ based on stellar population measurements is estimated to be $D=(16.8\pm0.8)~\textrm{Mpc}$ \cite{Blakeslee:2009tc,Bird:2010rd,Cantiello:2018ffy}, while the masses, $M\sim 6.2^{+1.1}_{-0.5}\times 10^9 M_\odot$ \cite{Gebhardt:2011yw} and $M\sim 3.5^{+0.9}_{-0.3}\times 10^9 M_\odot$ \cite{Walsh:2013uua}, are based on stellar dynamics and gas dynamics studies, respectively. On the other hand, the mass of M87* reported by the EHT Collaboration is $M=(6.5\pm 0.7)\times 10^{9}~M_{\odot}$. Note that, the mass estimation by the EHT collaboration is performed using \gr\ as the background model and hence is biased towards the same.

\begin{figure}[htb]
\begin{center}
\includegraphics[scale=0.6]{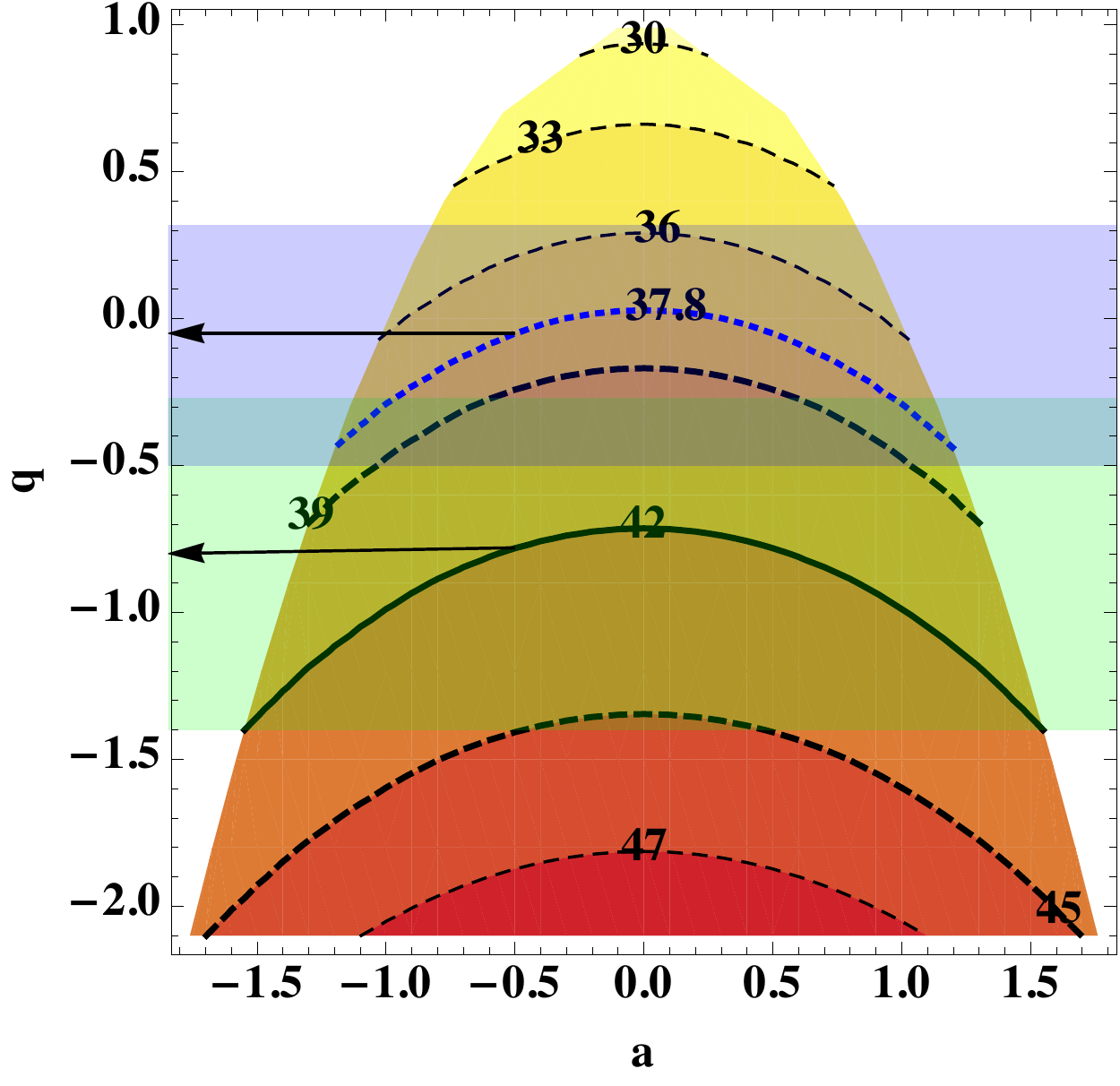}
\hfill
\includegraphics[scale=0.6]{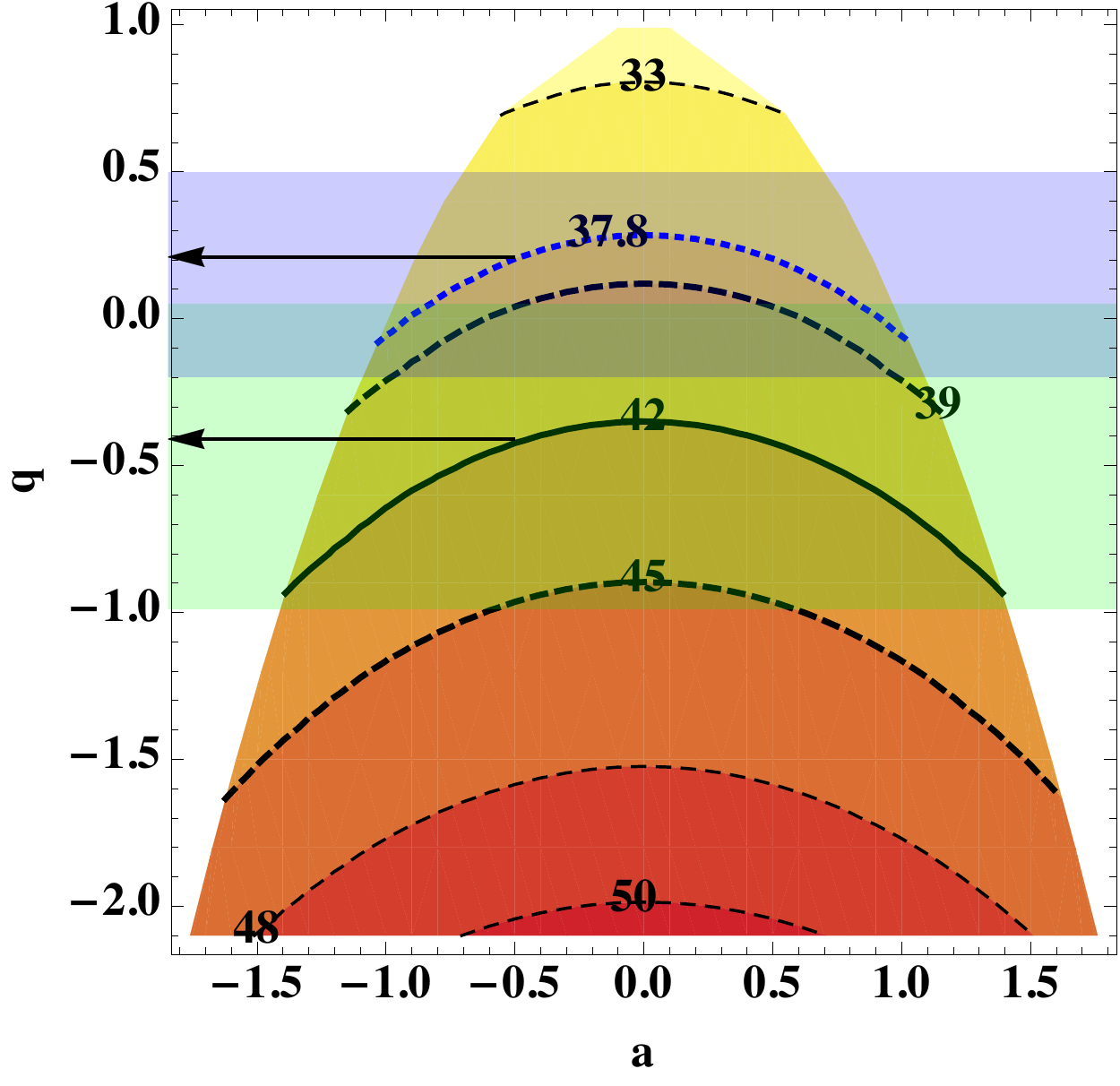}
\caption{The variation of the angular diameter of M87* in the (a,q) plane has been presented with (a) $M=6.2\times 10^9 M_\odot$ and $D=16.8~\textrm{Mpc}$ (the left figure) and (b) $M=6.5\times 10^9 M_\odot$ and $D=16.8~\textrm{Mpc}$ (the right figure). As both the figures explicitly demonstrate, the best estimate for the angular diameter of the shadow, namely $42~\mu\textrm{as}$ along with any possible error associated with it (shown by green band) is consistent with negative choices of the tidal charge parameter. Furthermore, even if we take into account a possible 10\% offset in the measurement of the angular diameter of the shadow, the negative tidal charge parameter remains well within the associated error estimations (as the blue band explicitly shows).}\label{Fig_AD_M87}
\end{center}
\end{figure}

In \ref{Fig_AD_M87} we plot the angular diameter of M87* based on the above distance measurement and mass estimates from stellar dynamics and EHT collaboration. To estimate the tidal charge parameter associated with a certain angular diameter we need to know an estimation for the rotation parameter, which according to EHT collaboration, lies within the range $0.5\lesssim |a|\lesssim 0.94$. Thus for a conservative estimate, we choose $|a|=0.5$, which along with the reported angular diameter measurement of the black hole shadow, i.e., $42~\mu\textrm{as}$ yields $q\sim-0.8$ for the left figure and $q\sim -0.4$ for the right figure in \ref{Fig_AD_M87}. Besides the fact that the tidal charge associated with the central value is \emph{negative} in both the contexts, it is also evident from \ref{Fig_AD_M87}, that inclusion of the associated error $\sim \pm 3~\mu\textrm{as}$ (depicted by the green band in \ref{Fig_AD_M87}) also favours negative values of the tidal charge. If we take the mass of M87* to be that obtained by stellar dynamics, then even if the 10\% offset of the shadow diameter is taken into account, which reduces the angular diameter to $37.8~\mu\textrm{as}$, the associated tidal charge parameter is negative ($q\sim-0.07$). Here also negative $q$ values are within 68\% confidence level (shown by the blue band in \ref{Fig_AD_M87}). While for the mass estimation of the EHT collaboration, the 10\% offset makes the tidal charge positive at the central value, but the negative tidal charges are very much within the 68\% confidence level. It is worth emphasizing that the mass of M87* estimated by the EHT Collaboration ($M=(6.5\pm 0.7)\times 10^{9}~M_{\odot}$ \cite{Akiyama:2019cqa,Akiyama:2019fyp,Akiyama:2019eap}) is performed assuming the background geometry to be Kerr, i.e., a general relativistic solution.
 
We emphasize that even if the 10\% offset \cite{Akiyama:2019eap} between the shadow region and the emitting region which surrounds it is taken into account, a negative tidal charge parameter can explain all the findings with all the independent mass measurements. Further, in most of the cases it is the negative tidal charge, which occupies most of the region within the 68\% confidence level (see \ref{Fig_AD_M87}). We further note that the above results are based on the choice of the rotation parameter being $\sim 0.5$. However, a recent analysis \cite{Tamburini:2019vrf} predicts the rotation parameter to be $|a|\sim 0.9$. In which case, as evident from \ref{Fig_AD_M87}, negative tidal charge parameter will be the most favourable scenario irrespective of the mass estimation of the M87*. This further boosts our result.

\begin{figure}[H]
\begin{center}
\subfloat[Estimation for the deviation from circularity, $\Delta C$, for inclination angle of $17^{\circ}$ associated with the supermassive black hole M87*.]
{\includegraphics[scale=0.5]{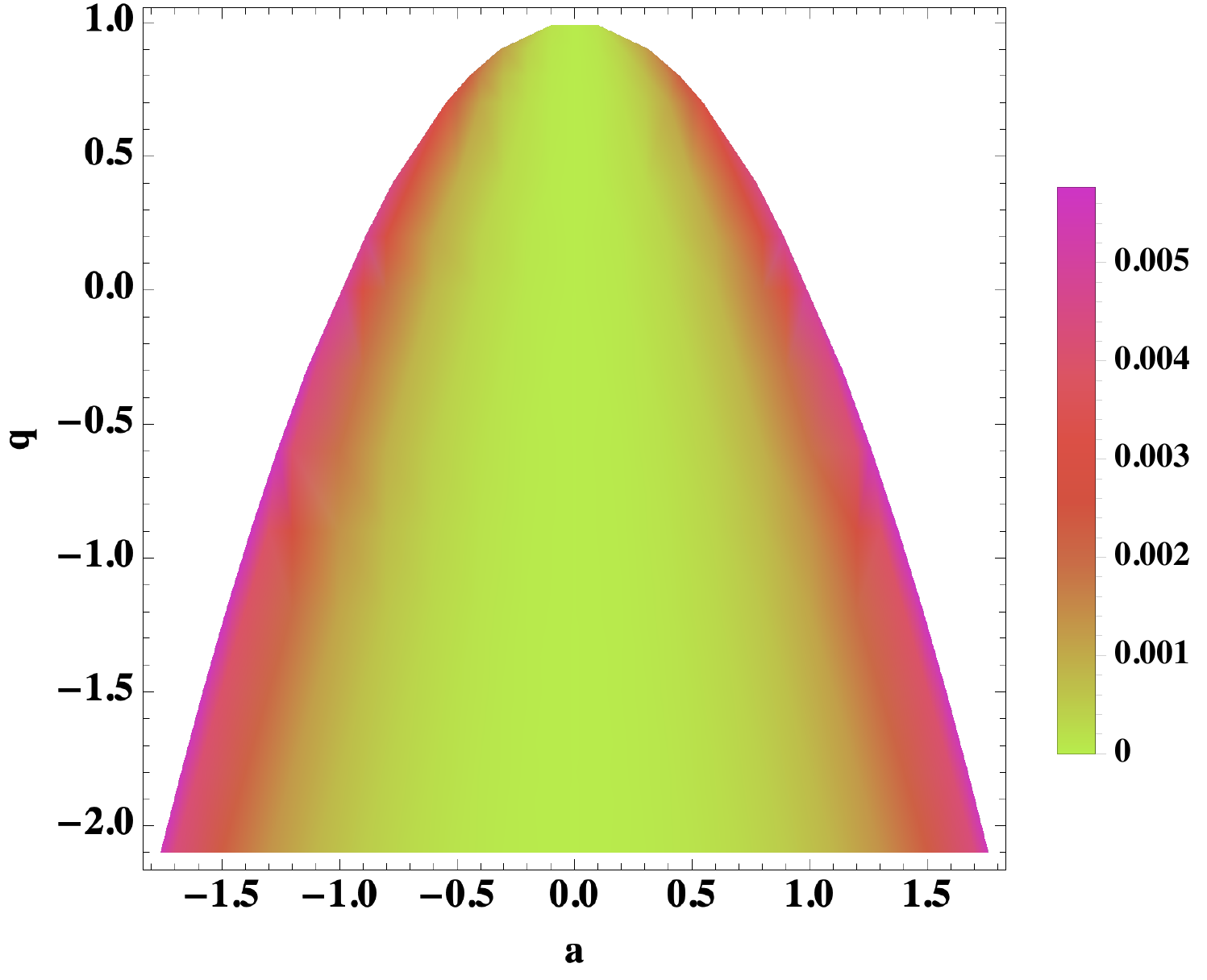}}
\hfill
\subfloat[Axis ratio $(\Delta \beta/\Delta \alpha)$ for inclination angle of $17^{\circ}$ associated with the supermassive black hole at the centre of M87 has been depicted.]
{\includegraphics[scale=0.5]{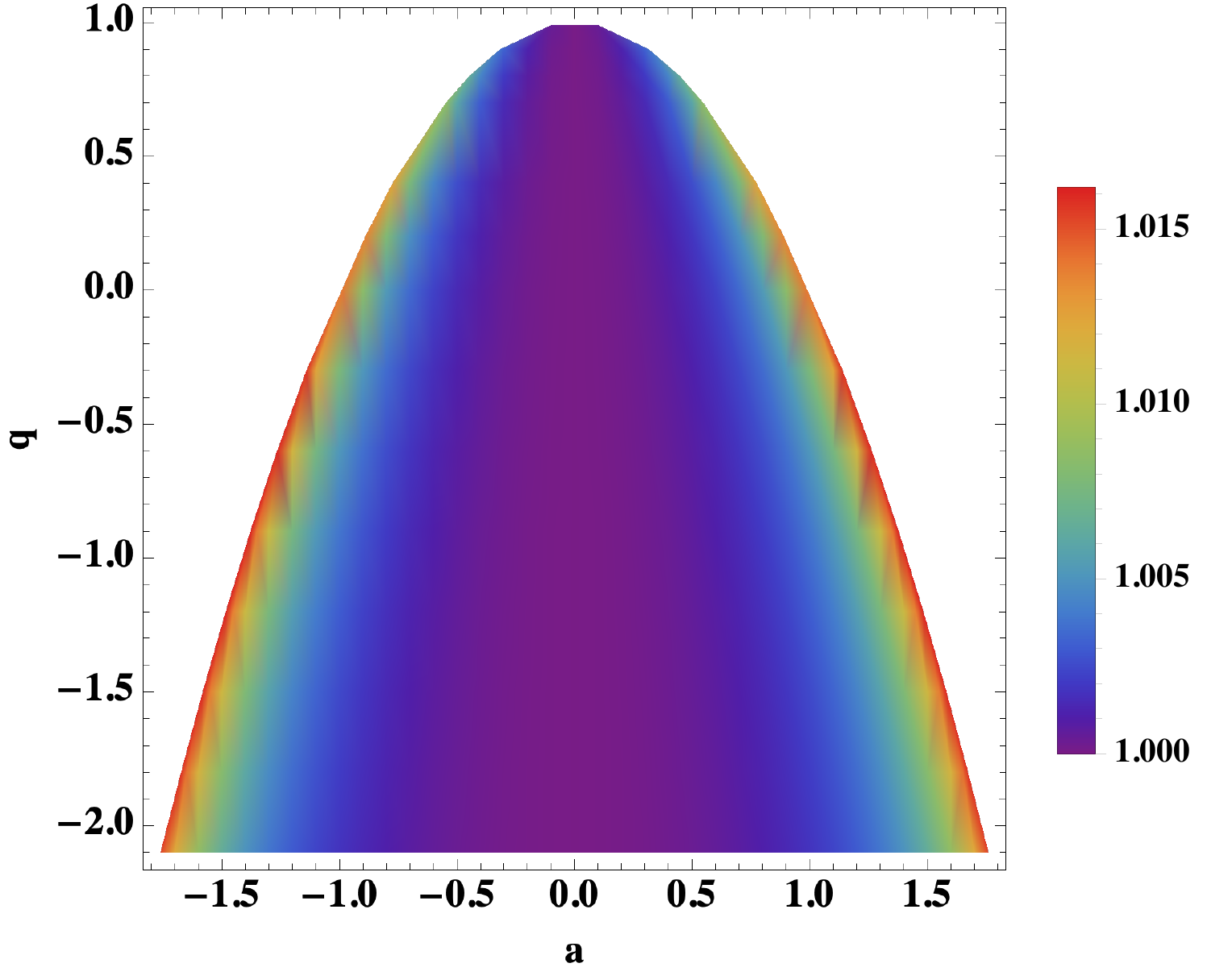}}
\caption{The deviation from circularity and axis ratio have been plotted in the $(a,q)$ plane using the parameters of the supermassive black hole at the centre of M87, provided by the EHT collaboration.}\label{Fig_DeltaC_M87}
\end{center}
\end{figure}

Finally, from \ref{papereq2} and expressions for $\alpha$ and $\beta$ it follows that, $\Delta C\sim 0.1\%$ and $\Delta A\sim 1.106$, even for extreme situations. This can be attributed to the very small inclination angle of M87*. While these are within the limits reported by the EHT Collaboration ($\Delta C<10\%$ and $\Delta A <4/3$ \cite{Akiyama:2019cqa}), no further constraints on $q$ can be derived from these observables. Thus the parameter space derived using the shadow diameter is consistent with the bounds on deviation from circularity and the axis ratio. This has been demonstrated explicitly in \ref{Fig_DeltaC_M87}.

\section{Concluding Remarks}

In this letter, we comment on the possible existence of extra dimensions from the measurement of shadow of M87* by the EHT Collaboration. We consider rotating braneworld BHs whose distinctive feature is the existence of a \emph{negative} tidal charge $q$. We compute the angular diameter of the shadow of M87* from its known mass, distance and inclination and try to determine the tidal charge from the measurement of angular diameter of M87*. Choosing a conservative estimate for the rotation parameter, namely $|a|\sim0.5$, we see that the observed angular diameter of $42~\mu\textrm{as}$ with mass measurement from stellar dynamics report a \emph{negative} tidal charge parameter. The result holds true even if we take into account a 10\% offset in the angular diameter of the shadow. Further, the negative tidal charge parameter remains more favoured even if the error associated with angular diameter measurement is taken into account. For the mass measurement from EHT collaboration itself, the angular diameter of $42~\mu\textrm{as}$ predicts a negative tidal charge, while with the 10\% offset, the charge can be come positive. Still, the negative values of the tidal charge are well within the error bounds. Further, if the spin of the black hole is considered to be higher, then also the negative tidal charges will be more favoured. We emphasize that the case of vanishing tidal charge (i.e., \gr) is definitely within the error limit of $\pm 3~\mu\textrm{as}$ in most of the cases and the shadow measurement by no means rule out \gr. However, our analysis reveals that the negative values of the tidal charge parameter seems to be more favoured in the sense described above. Interestingly, this turns out to be consistent with our earlier finding from the study of electromagnetic emission from the accretion disk around quasars \cite{Banerjee:2017hzw,Banerjee:2019sae}.

From the theoretical side, given the above preference for negative values of the tidal charge parameter, following \cite{Chamblin:2000ra}, we can argue that the extra dimension must be compactified. This is because it is this sign and magnitude of the tidal charge, that determines the extent of the extra dimension as well as the penetration of the black hole horizon into the bulk spacetime \cite{Chamblin:2000ra}. In particular, negative values of the tidal charge shrinks the extent of the extra dimension. Qualitatively, following \cite{Chamblin:2000ra}, one may argue that for $q=-0.5$, which is a reasonable estimation of the tidal charge from the black hole shadow measurement, reduces the extent of extra dimension, in comparison to the smaller values of $q$, by almost 5\%. Hence with sufficient number of shadow measurements, one should be able to determine the tidal charge parameter and thus comment on the compactification scale associated with the extra dimension. Presence of a negative tidal charge will also hint towards a quantum nature of the black holes from the AdS/CFT perspective \cite{Emparan:2002px}. This makes any observational signature for negative tidal charge parameter $q$, as is the case here, important from various theoretical perspectives as well. More detailed theoretical implications of our results will be presented elsewhere.
 
\section*{acknowledgements}

The research of SSG is partially supported by the Science and Engineering Research Board-Extra Mural Research Grant No. (EMR/2017/001372), Government of India. Research of S.C. is funded by the INSPIRE Faculty Fellowship (Reg. No. DST/INSPIRE/04/2018/000893) from the Department of Science and Technology, Government of India.

\bibliography{Black_Hole_Shadow,Brane,KN-ED}

\bibliographystyle{./utphys1}
\end{document}